\documentstyle[aps,twocolumn,epsfig]{revtex}

\begin{document}

\title{\bf Marcatili's Lossless Tapers and Bends: an Apparent Paradox and
its Solution}

\author{Antonio-D. Capobianco$^1$, Stefano Corrias$^1$, Stefano Curtarolo$^{1,2}$ 
and Carlo G. Someda$^{1,3}$ \\
$^1$DEI, Universit\`a di Padova, Via Gradenigo 6/A 35131 Padova, Italy. \\
$^2$Dept. of Materials Science and Engineering, MIT, Cambridge, MA 02139, USA\\
$^3${corresponding author e-mail: someda@dei.unipd.it}
\begin{center}
{\sf Proceedings of Jordan International Electrical and Electronic Engineering Conference, \\
JIEEEC'98, April 27-29, 1998, Amman, Jordan}
\end{center}
}
%\date{}

\maketitle

\thispagestyle{empty}

\section{Abstract.}

Numerical results based on an extended BPM algorithm indicate that,
in Marcatili's lossless tapers and bends, through-flowing waves are
drastically different from standing waves. The source of this
surprising behavior is inherent in Maxwell's equations. Indeed, if
the magnetic field is correctly derived from the electric one, and
the Poynting vector is calculated, then the analytical results are
reconciled with the numerical ones. Similar considerations are
shown to apply to Gaussian beams in free space.

\section{Introduction.}

In 1985, Marcatili \cite{Mar85} infringed a historical taboo, by showing
that lossless tapers and bends in dielectric waveguides can be conceived
and designed, at least on paper. The key feature shared by all the
infinity of structures which obey Marcatili's recipe, is the fact that
the phase fronts of the guided modes which propagate in them, are
closed surfaces. As well known, phase fronts which extend to infinity
in one direction orthogonal to that of propagation do entail radiation
loss, but closed fronts can avoid this problem. However, shortly after 
the first recipe \cite{Mar85}, it was pointed out \cite{MaSo87} that
that recipe could generate
some inconsistencies. In fact, a traveling wave with a closed phase
front is either exploding from a point (or a line, or a surface), or
collapsing into such a set. In a lossless medium where there are no
sources, this is untenable. On the other hand, it was also pointed
out in \cite{MaSo87} that a standing wave with closed constant-amplitude
surfaces is physically meaningful. Therefore, propagation of a
through-flowing wave through any of Marcatili's lossless tapers or bends
has to be described in this way: the incoming wave must be decomposed
as the sum of two standing waves, of opposite parity with respect to
a suitable symmetry surface. The output wave was then to be found
as the sum of the values taken by the two standing waves at the
other end of the device. Another point raised in \cite{MaSo87} was that
very similar remarks apply to Gaussian beams in free space.

Later on, the literature showed that interest in this problem was not
so high, for a long time. Recently, though, we observed several symptoms
of a renewed interest in low-loss \cite{MKM94,Man95,Vas94} and lossless
\cite{Wu96,Lee97} tapers or bends. This induced us to try to go beyond the
results of \cite{MaSo87}, and to clarify further the difference between
through-flowing and standing waves in Marcatili's tapers.

The new results reported in this paper can be summarized
as follows. In Section~III, we
show that the numerical analysis (based on an extended BPM algorithm)
of Marcatili's tapers reconfirms that indeed through-flowing waves are
drastically different from standing ones. The latter ones match very
well the analytical predictions of the original recipe \cite{Mar85}, but
through-flowing waves have open wave fronts, which do not entail any
physical paradox. In Section~IV, we provide an analytical discussion
of why, in contrast to what occurs with plane waves
in a homogeneous medium and with guided modes in longitudinally
uniform waveguides, through-flowing waves are so different from standing
ones. We show that this is a rather straightforward consequence of
Maxwell's equations. From this we will draw the conclusion that
a through-flowing wave propagating in one of Marcatili's tapers is
never strictly lossless. Nonetheless, our numerical results reconfirm
that the recipes given in \cite{Mar85} do yield extremely low radiation
losses.

Finally, we address briefly the case of Gaussian beams in free space, and
explain why they behave essentially in the same way
as the devices we discussed above. In fact, Maxwell's equations show
that in general the phase fronts of the magnetic field in a Gaussian
beam are not the same as the phase fronts of the electric field.
Therefore, the Poynting vector is not trivially proportional to the
square of the electric field. Consequently, a through-flowing beam,
resulting from two superimposed standing waves of opposite parities,
can be surprisingly different from the parent waves.

\section{Numerical results.}

The geometry of Marcatili's tapers can eventually be very complicated
(e.g., see \cite{SaMa-1991}). For our tests, however, we chose
a simple shape, to avoid the danger that geometrical features could hide
the basic physics we were trying to clarify. The results reported here
refer to a single-mode taper whose graded-index core region is delimited
by the two branches of a hyperbola (labeled A and A' in Figs.~1 and 2), and
has a mirror symmetry with respect to its waist. This is a ``superlinear''
taper, according to the terminology of \cite{Mar85}, with an index
distribution (see again \cite{Mar85})

\begin{equation}
n=\left\{
\begin{tabular}{ll}
$n_0\sqrt{1+2\Delta/(cosh^2\eta-sin^2\theta)}$ & $\theta_1<\theta<\theta_2$ \\
$n_0$ & $\theta_1>\theta>\theta_2$
\end{tabular}
\right.
\end{equation}

where $\eta$ and $\vartheta$ are the elliptical coordinates, in the plane
of Figs.~1 and 2. Fig.~1 refers to a standing wave of even 
symmetry with respect to the waist plane, Fig.~2 to a standing wave
of odd symmetry. The closed lines
are constant-amplitude plots. We see that they are essentially elliptical,
so they agree very well with the predictions of \cite{Mar85}.

\begin{figure}[h]
%\label{fig1}
%\vspace{9cm} 
\centerline{\epsfig{file=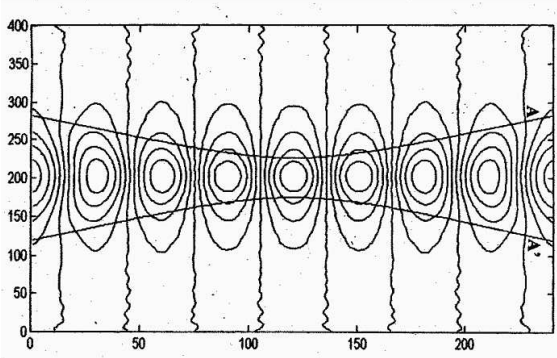,height=7.5cm,angle=-90}}
 \vspace*{3mm}
\caption{Constant-amplitude plot of a standing wave of even symmetry
(with respect to the waist plane) in a superlinear
Marcatili's taper.}
\end{figure}

As mentioned
briefly in the Introduction, these results were generated using an
extended BPM, which deserves a short description. In fact, it is
well known that the standard BPM codes are suitable to track only
traveling waves, as they neglect backward waves. Our code (using a Pade's
operator of order (5,5)) also
generates a traveling wave, and its direction of propagation is inverted
whenever the wave reaches one of the taper ends. In order to
generate single-mode standing waves, each reflection should take
place on a surface whose
shape matches exactly that of the wave front. This is very difficult
to implement numerically, but the problem can be circumvented,
letting each reflection take place on a {\it phase-conjugation} flat
mirror. Our code calculates then, at each point in the taper, the sum of
the forward and backward fields, and stops when the difference
between two iterations is below a given threshold.

\begin{figure}[h]
%\vspace{9cm}
\centerline{\epsfig{file=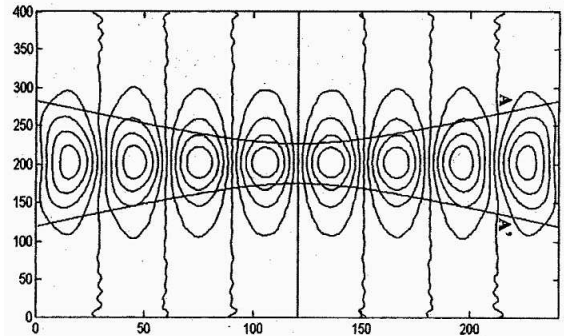,height=7.5cm,angle=-90}}
 \vspace*{3mm}
\caption{Constant-amplitude plot of a standing wave of odd symmetry
(with respect to the waist plane) in a superlinear
Marcatili's taper.}
\end{figure}

Figs.~3 and 4 refer to a through-flowing wave. The almost horizontal dark lines
in Fig.~3 are its phase fronts. They are drastically different from those
predicted by the analytical theory in \cite{Mar85}, which are
exemplified in the same figure as a set of confocal ellipses. Note that
the through-flowing wave has been studied numerically in two ways. One was simply
to launch a suitable transverse field distribution, and track it down the
taper with a standard BPM code. The other one was to calculate the linear
combination
(with coefficients $1$ and $j$) of the even and odd standing waves
shown in Figs.~1 and 2. The results obtained in these two ways were
indistinguishable one from the other. This proves that indeed through-flowing
waves are drastically different from standing ones. In particular,
as we said in the Introduction, they keep clear from any
paradox connected with energy conservation.

Fig.~4 shows a field amplitude contour plot for the same
through-flowing wave as in Fig.~3. It indicates that propagation through the
taper is indeed almost adiabatic.
Therefore, as anticipated in the Introduction, insertion losses of
Marcatili's tapers are very low (at least as long as the length to
width ratio is not too small), although they are not strictly zero. As a
typical example, for a total taper length of $2.5 \mu m$, a waist
width of $0.55 \mu m$ and an initial-final width of $1.65 \mu m$,
BPM calculations yield that the lost power fraction
is $1.4 \times 10^{-4}$. A typical plot of power vs. distance
along a taper with these features is shown in Fig.~5.

\begin{figure}[h]
%\vspace{9cm}
\centerline{\epsfig{file=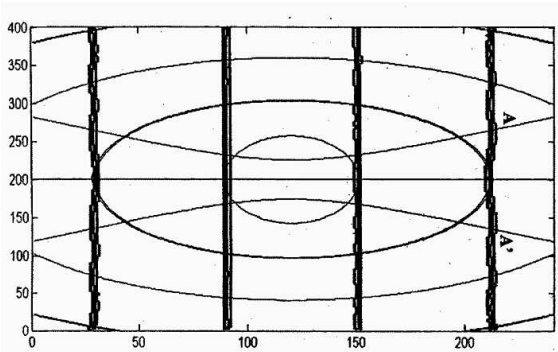,height=7.5cm,angle=-90}}
 \vspace*{3mm}
\caption{Phase fronts of a through-flowing wave in the same superlinear
taper as in Figs. 1 and 2.}
\end{figure}

\section{Theoretical discussion.}

For the sake of clarity, let us restrict ourselves to the case of
two-dimensional tapers, like those of the previous section, where
the geometry and the index distribution are independent of the
$z$~coordinate, orthogonal to the plane of the figures. However,
our conclusions will apply to 3-D structure also.

The index distributions found in the corner-stone paper \cite{Mar85} are such
that the TE modes (electric field parallel to $z$) satisfy rigorously a wave
equation which can be solved by separation of variables. Obviously,
the same equation is satisfied rigorously by the transverse component of
the magnetic field. 

However, in general two solutions of these two
wave equations which are identical, except for a proportionality
constant, {\ do not satisfy Maxwell's equations} in that structure.
This is very easy to show, for example, for the case which was called
``linear taper'' in \cite{Mar85}, namely, a wedged-shape region with
a suitable index distribution, where a guided mode propagates in the
radial direction. The claim \cite{Mar85} that the dependence of
$E_z$ on the radial coordinate is expressed by a Hankel function of
imaginary order $i\nu$, related to other features of the taper, is
perfectly legitimate. However, one cannot extrapolate from it that the
same is true for the magnetic field. In fact, calculating the curl of the
electric field we find that the azimuthal component of the magnetic field
is proportional to the {\it first derivative} of the Hankel function,
which is never proportional to the function itself. The same is true
for the Mathieu function of the fourth kind, which satisfy the
wave equation in the coordinate system which fits the superlinear
taper of the previous Section. This entails a drastic difference with
plane waves, and with guided modes in uniform waveguides, where the
derivative of the exponential function that describes the propagation of
the electric field is proportional to the function itself. In the cases
at hand, the concept of {\it wave impedance} becomes ill-grounded.
In fact, the electric field and the transverse magnetic field have
identical dependencies on the transverse coordinate, so that their
ratio is constant over each wavefront, but they are different functions
of the {\it longitudinal} coordinate, as if the `wave impedance'
were not constant at all along the wave path. This indicates why
it is very risky, in the case at hand, to make general claims on the
Poynting vector starting from the spatial distribution of
only the electric field.
To strengthen our point, let us prove explicitly that it is not
self-consistent to claim that a {\it purely traveling} TE wave, whose radial
dependence is expressed by a Hankel function of imaginary order, $H_{i\nu}$,
can propagate along a linear taper. As we just said, for such a wave
$E_z$ is proportional to $H_{i\nu}$, $H_\psi$ is proportional to
$H'_{i\nu}$, so the radial component of the Poynting vector is
proportional to $H_{i\nu} (H'_{i\nu})^*$. In a purely traveling wave
there is no reactive power in the direction of propagation. Combining
with what we just said, it is easy to see that this would imply
$|H_{i\nu}|^2=constant$ along the (radial) direction of propagation,
a requirement that is not satisfied by Hankel functions. (Note, once more,
that it is satisfied by exponential functions). Therefore, {\it any}
wave along a linear taper whose radial dependence is expressed as a
Hankel function must be at least a partially standing wave. A
through-flowing wave, if it exists, must behave in a different way.

\begin{figure}[ht]
%\vspace{9cm}
\centerline{\epsfig{file=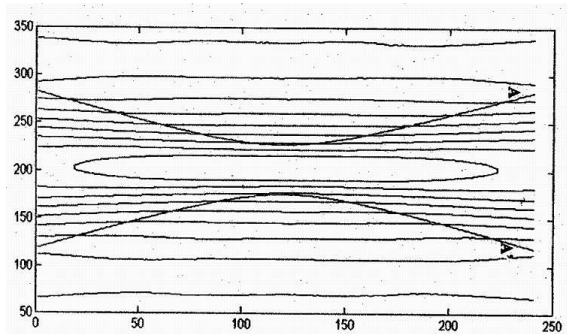,height=7.5cm,angle=-90}}
 \vspace*{3mm}
\caption{Field-amplitude contour plot, in the same superlinear
taper as in the previous figures.}
\end{figure}

Finally, let us address briefly the case of Gaussian beams in free space.
It was pointed out in \cite{MaSo87} that they behave essentially
in the same way as the devices we discussed above. There is still
something to add to the discussion of \cite{MaSo87}.  
Assume that the electric field of an electromagnetic wave has the
classical features of a $TEM_{00}$ Gaussian beam (see, e.g., \cite{Som98}).
Then, Maxwell's equations show that the phase fronts of the magnetic
field are not the same as those of the electric field, neither on the
waist plane nor far from it. Hence, the Poynting vector is not
trivially proportional to the square of the electric field. This entails
the presence of a reactive power (never accounted for in the classical
classroom explanations of Gaussian beams), and an active power flow
which is not always along the lines orthogonal to the electric
field phase fronts. Once again, a through-flowing beam,
resulting from two superimposed standing waves of opposite parities,
is different from the parent waves, and the difference is maximum
on the symmetry plane, i.e. at the beam waist. Due to time and space
limits, the details of this discussion must be left out of this
presentation, and will be published elsewhere \cite{CaSo}.

\begin{figure}[h]
%\vspace{9cm}
\centerline{\epsfig{file=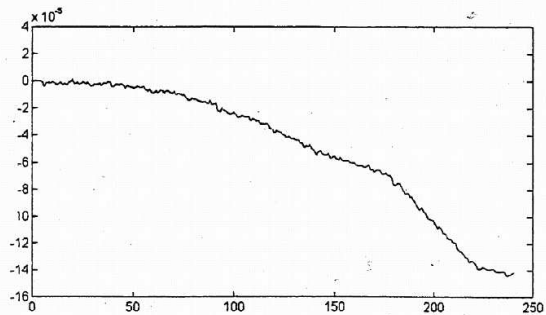,height=4.5cm}}
 \vspace*{3mm}
\caption{Power vs. distance, in a superlinear taper of the shape
shown in the previous figures, whose parameters are specified in
the text.}
\end{figure}

\section{Conclusion.}

We tried to shed new light on an old problem, namely, whether the
idea of a guided mode traveling without any loss through a dielectric
taper can be sustained without running into any physical paradox.
Our numerical results, obtained with an extended BPM technique, have
fully reconfirmed what was stated in \cite{MaSo87}: in Marcatili's tapers,
standing waves have the basic features outlined in \cite{Mar85}, but
through-flowing waves do not. This prevents them from running into a
paradox, but on the other hand entails some loss, although very small
indeed. Next, we have provided an explanation for the unexpected and
puzzling result, a drastic difference between standing and through-flowing
waves in the same structures. The source of these ``surprise'' is
within Maxwell's equations.

It was pointed out in \cite{MaSo87} that some of the problems discussed
here with reference to Marcatili's tapers apply to Gaussian beams
in free space as well.

\end{document}